\documentclass[conference]{IEEEtran}
\usepackage[dvips]{graphicx}

\usepackage{amsmath}
\usepackage{amsfonts}
\usepackage{amssymb}
\usepackage{algorithm}
\usepackage{algorithmic}
\ifCLASSINFOpdf
   \usepackage[pdftex]{graphicx}
\else
   \usepackage[dvips]{graphicx}
   \graphicspath{{./eps/}{./jpeg}}
   \DeclareGraphicsExtensions{.eps,.pdf}
\fi

\begin{document}
%
% paper title
% can use linebreaks \\ within to get better formatting as desired
\title{Zero Energy Network stack for Energy Harvested WSNs}

% author names and affiliations
% use a multiple column layout for up to three different
% affiliations
\author{\IEEEauthorblockN{ Akshay Uttama Nambi S. N$^1$, Prabhakar T.V$^2$, R Venkatesha Prasad$^1$, Jamadagni H.S$^2$}
\IEEEauthorblockA{
$^1$Delft University of Technology, The Netherlands\\
$^2$Department of Electronic Systems Engineering (DESE), Indian Institute of Science, Bangalore, India\\
\{Akshay.narashiman,R.R.Vekateshaprasad\}@tudelft.nl
\{tvprabs,hsjam\}@dese.iisc.ernet.in
}}

\maketitle

\begin{abstract}

We present our ``Zero Energy Network'' (ZEN) protocol stack for energy harvesting wireless sensor networks applications. The novelty in our work is $4$ fold: (1) Energy harvesting aware fully featured MAC layer.  Carrier sensing, Backoff algorithms, ARQ, RTS/CTS mechanisms, Adaptive Duty Cycling are either auto configurable or available as tunable parameters to match the available energy (b) Energy harvesting aware Routing Protocol. The multi-hop network establishes routes to the base station using a modified version of AODVjr routing protocol assisted by energy predictions. (c) Application of a time series called ``Holt-Winters'' for predicting the incoming energy. (d) A distributed smart application running over the ZEN stack which utilizes a multi parameter optimized perturbation technique to optimally use the available energy. The application is capable of programming the ZEN stack in an energy efficient manner. The energy harvested distributed smart application runs on a realistic solar energy trace with a three year seasonality database. We implement a smart application, capable of modifying itself to suit its own as well as the network's energy level. Our analytical results show a close match with the measurements conducted over EHWSN testbed. 
\end{abstract}

\IEEEpeerreviewmaketitle

\section{Introduction}

Several disruptive sensing and control algorithms proposed for wireless sensor networks (WSNs) have remained abeyant due to power requirements for the hardware. However, recently Energy Harvesting wireless Sensor network (EHS) applications are a reality due to their independence from utility power and thus unleashing several interesting proposals in sensing and control. This possibility is because of advancements in nanoelectronics and materials, increased power efficiency of harvesting electronics and the rapid advancements in high integrated ultra low power microcontrollers and communication radios.  The ``ZigBee Green'' \cite{ZigBee_Green} is specifically meant for running out of energy harvesting sources.  Recent advancements in materials and MEMS research has made Thermo Energy Generators (TEGs) and vibration harvesters (unusable till recently) as potential energy sources. Other contributing parameters include system operating voltage \& frequency of the microcontroller and radio, and finally the system's several low power modes.  Operating voltages of about 1.8 volts with operating frequencies of less than $1$MHz and sleep currents of the order of $100$ nA are some of the recent technology trends for system parameters.  Radio communication energies of about $10$ nJ per bit \cite{ansem} is fast becoming a reality. Also, energy storage in thin film batteries \cite{cymbet}\cite{ST_micro} and low leakage super capacitors \cite{nesscap} \cite{capxx}  with several thousand charge-discharge cycles offer efficient energy storage.

EHS find attractive applications wherever remote monitoring and control are required and cover a wide spectrum of scenarios. On the one end could lie an intrusion detection system deployed in a wireless tripwire paradigm for monitoring an international border, and the other end of the spectrum is a simple wireless switch application for home lighting systems. While the former application requires an energy storage buffer, the latter application requires energy generation and its usage on the fly.  Also, these applications work in multi-hop and single-hop settings respectively. Other applications under this wide spectrum include intelligent transportation, smart buildings, pollution monitoring, agriculture and climate change, health care including body area networks and other similar applications.  For outdoor applications, while photovoltaic panels offer significantly higher power compared to other harvesting sources, often sensor nodes packaged with these small panels could physically be placed where reflected light or even a partial shade might be present.  This is especially true for intrusion detection where the purpose is to detect an intruder under a camouflage of the sensor node so as to avoid attention from the intruder. Additionally, seasonal variation in sunlight plays an important role in ensuring continuous and untethered operation of solar powered EHS nodes. Thus, applications perhaps have to continuously adjust to varying instantaneous power fluctuations and yet accomplish their primary assigned task. Since the EHS nodes are usually wide spread and the communication of these nodes are limited in range, multi-hop network is required with performance deterioration within acceptable limits. Thus, designing a multi-hop EHS sensor network under harsh incoming energy condition is indeed a challenge.

One key difference between battery driven WSNs and EHS networks concerns the optimization parameters such as: (a) energy consumption to increase the node's battery life and (b) A network wide policy such as support a network lifetime of about ``X'' (say) number of hours or support the largest partitioned network. A large body of work in WSN is limited to maximizing the policy subject to a given energy constraint or minimizing energy consumption to satisfy a network policy requirement. In direct contrast, for an energy harvested network, the objective is to maximize the network policies and also maximize the energy consumption. In other words, minimize the energy differential between the available and consumed energy in each discrete time slot and maximize the network policy. Our work in this paper uses a multi-criteria optimization approach to find the optimal network policy and energy differential curve. The significance of optimal network policy-energy differential curve is to show a trade-off between global perspective of network policy and energy differential. Also, when the requirement on either network policy or energy consumption varies, one can use the optimal network policy-energy differential curve to locate the new optimal trade-off.

\begin{figure}[ht]
\centering
\centering \includegraphics[scale=0.38]{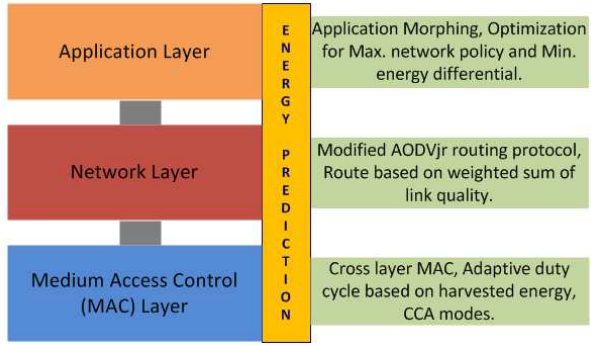}
\caption{ Energy Harvesting Sensor node stack. }
\label{stack}
\end{figure}

In this paper, the goal is to implement a distributed smart application for environment monitoring applications where the sensor and communication nodes are powered with harvested energy. We approach this goal by building a protocol stack called the ``ZEN Stack'' as shown in Fig. \ref{stack}. At the application layer, we consider the application performance not only its own energy level, but also on its neighbour and parent forwarding nodes. We use energy measurements and time series predictions to change the behaviour of our base application in a manner that energy utilization is a maximum in the time slot.  At the routing layer, since each node in the network can harvest different energy magnitudes and thus have a varying energy profile, finding network wide routes when energy levels on nodes vary continuously is a challenge. Hence, routing protocols have to consider the energy harvested to decide how to route packets to the base station. At the data link layer, we propose several enhancements to the existing Medium Access Control (MAC) Layer. Our goal is to study the performance of this stack in its ability to exploit the available energy to the maximum extent possible.  Thus ensuring network wide ``energy neutrality'' condition.  The analysis followed by implementation results are encouraging and also show that even ``near real time'' applications can perform reasonably well.

%In this paper, the goal is to implement a multi-hop multi-node Energy Harvesting WSN (EHWSN). In a multi-hop EHWSN, not just the application layer has to be harvesting aware (adjusting the application rate based on harvested energy?) but also lower Routing and Medium Access Control (MAC) layers have to consider harvested energy for efficient route selection and duty cycling. To achieve our goal, we propose enhancements to the existing routing and MAC protocols for EHWSN. Since each node in the network can harvest different energy magnitudes and thus have a varying energy profile, finding network wide routes when energy levels on nodes vary continuously is a challenge. Hence, routing protocols have to consider the energy harvested to decide how to route packets to the base station. At the data link layer, we propose several enhancements to the existing MAC Layer where we enable virtual energy transfer between peer nodes in the network based on the harvested energy. We assist the routing and MAC protocols with future energy predictions using time series energy prediction models. 

\section{Related work}

In recent times, a large body of work is being concentrated on Energy Harvesting Wireless Sensor Nodes (EHWSN). Several energy harvesting sources like solar, thermo, acoustic, wind, RF and wave for driving low power embedded devices are discussed in \cite{surveyehs}. Many energy harvesting WSNs have been implemented in past like Trio \cite{trio}, AmbiMax \cite{Ambimax}, Prometheus \cite{prometheus}. However, literature on Energy Harvesting systems is mostly limited to single node networks or extensive simulation study. Many of the work focus on the system building, efficiency and viability of energy harvesting mechanisms. In \cite{prometheus} authors shows improvement in lifetime where a supercapacitor and battery are charged together from a solar panel by adjusting their duty cycles. Recently Power management and data rate maximization in EHWSN was discussed in \cite{chandra}. The authors analytically find the required amount of energy to be harvested to ensure energy neutral operation in each slot. 

Accurate prediction of the energy is an important factor for energy neutral operations. Time series based energy predictions are commonly used for predicting solar energy. Exponential Weighted  Moving Average (EWMA) is commonly applied for future energy predictions \cite{aman}, \cite{dynamic} and assigns weights to data in the time series, assigning lower weights to older data and giving importance to more recently acquired data. Holt-Winters (HW) time series is widely used in stock markets trend analysis, production planning, healthcare decisions on staffing \& purchasing, demand forecasting \cite{hw1} and for predicting share prices  where forecast value takes trend and seasonal variation into account \cite{hw}, \cite{hw2}. In \cite{infocomm_2011} a three state markov chain model is used to predict the solar energy source and their transition probabilities are restricted for day time only. Energy samples from a simulation model are also generated from the model.

In the network layer, the work in \cite{dynamic} demonstrate the performance benefits in using AODVjr for ZigBee mesh networks. The work introduces an energy aware metric with future energy consumption using Exponential Weighted Moving Average (EWMA) time series model. The work is limited to simulation studies and also considers battery driven sensor nodes. In this work, we utilize AODVjr protocol \cite{aodvjr} a simplified version of AODV routing protocol that reduces implementation complexity by eliminating certain features from the original AODV protocol. The work in \cite{Hasenfratz} compares several energy harvesting based routing protocols in a simulation setup and show that R-MPRT routing protocol outperforms the other protocols. The work considers realistic MAC protocols suitable for energy harvesting networks and is limited to extensive simulation studies. In \cite{energyaware}, authors propose a routing algorithm that takes into account distance and link quality information. The algorithm requires that all nodes in the network know about the position, which usually not available in sensor network deployments. The authors in \cite{E-wme}, propose asymptotically Optimal Energy-Aware Routing for Multihop Wireless Networks named Energy-opportunistic Weighted Minimum Energy (E-WME). In E-WME the best route is decided based on the energy considerations, however, routing decisions should also take into account different channel conditions, especially in a wireless environment in order to minimize packet retransmissions and energy wastage. 

At the data link layer, while there are several MAC protocols been designed for WSN, they are not optimized for energy harvesting sensor nodes. However there are quite a few MAC protocols designed for EHS, the work in \cite{adaptive} proposes dynamically adapting the duty cycle of a node by observing deviations in energy input from an estimated model to ensure energy-neutral operation. In \cite{control} the authors propose to use adaptive control theory to formulate a linear-quadratic optimal tracking problem for maximizing task performance and minimal duty cycle variations. In \cite{macschemes} the authors present analytical models and performance metrics for different MAC protocols. The work however is limited to simulation studies and over single-hop network. In \cite{sleepwakeup}, performance of various sleep and wakeup strategies based on channel state, battery state and environmental factors are analysed. The authors propose a optimal sleep wakeup strategy using game theory approach which provides trade-off between packet dropping and blocking probabilities. 

Several optimization techniques have been discussed in literature for battery driven multihop WSN's in the past where the objective is to maximize the network lifetime by minimizing energy consumption \cite{delayconst}, \cite{num}. With a view on utilizing the harvested energy in an optimized manner, adapting the performance of an application while respecting the limited and time-varying amount of available power in EHWSN is discussed in \cite{adappowermoser}. A formal model that actively adapts application parameters such as ``rate of sensing'' is used to optimize the performance. The simulation study does not consider network related parameters or network wide energy levels.  A ``Lazy Scheduling Algorithm'' is proposed and tested for its effectiveness by assigning several power values to the system in \cite{moser}. Task admittance and future energy prediction is carried out with energy variability characterization curves generated by modeling the power source.

We demonstrate a distributed smart application which optimizes the necessary parameters to ensure maximization of application policy and minimization of residual energy. Contrasting to the work in literature, we believe performing Energy Neutral Operations based on node's own energy is not optimal. Instead we propose to derive the optimal number of operations and the residual energy based on the available energy on the node, predicted energy and network energy (i.e., neighbouring nodes). Particularly, we tune the duty cycle factor, transmission power factor and the application morphing factor to achieve optimal performance. Amongst the rich literature on EHWSN, \textit{inter alia}, we position our work based on these attributes: (a)~Harvested energy is the only available source; (b)~EHWSN nodes are deployed in a testbed with MAC layer contention and adaptive duty cycling; (c)~Inter-node distance of about $25-30$ft with a realistic channel where Wi-Fi and other 2.4GHz radios are present; (d)~Energy prediction from physical solar trace data including seasonality over a three year database is used; and (e)~A smart application which morphs itself into several forms based on the energy level of the node as well as its network in a manner where performance deterioration is within acceptable limits.

\section{Experimental Setup}

\begin{figure}[htp]
\centering \includegraphics[scale=0.26]{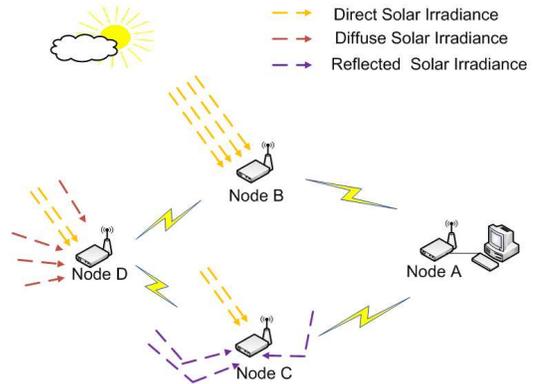}
\caption{Experimental Setup.}
\label{setup}
\end{figure}

Our multi-hop EHWSN comprises of $4$ nodes placed in a tree topology as shown in Fig \ref{setup}. Our custom hardware motes uses TI's MSP430 microcontroller and Chipcon's IEEE 802.15.4 standard CC2520 radio. We broadly divide the network into data collecting \textit{ sensor nodes }and data relaying \textit{relay nodes}. From Fig \ref{setup}, nodes B,C are relay nodes and node D is the sensor node. Relay nodes have the task of forwarding packets to the base station. Node A is utility powered and configured as the base station or data sink node for the network. For our implementation, we used solar energy harvesters with varying energy profile across the network. Since we require realistic spatio-temporal varying power across all the network nodes, the power output solar trace database available from \cite{lrss} is replicated over the experimental setup. The three year solar trace with trend and seasonality has power output for direct, diffused and reflected sunlight. In Fig \ref{setup}, Node B runs on direct sunlight profile, node C runs on reflected sunlight profile and node D runs on diffused sunlight profile. The power output from the solar panels was varied by switching ``on" and ``off" electrical lamps in the solar harvesters. We scaled down the power output obtained from Lowry Range Solar Station (LRSS) to match the laboratory solar harvesters.

Each node in the network has two modes of operation  namely (a) active mode and (b) sleep mode. While, in \textit{active mode}, a node can transmit and receive packets, in \textit{sleep mode}, node turns off the radio transceiver to reduce the energy consumption. Switching between these two modes depends on the sleep and wakeup strategy used by that node. In our setup, based on the harvested energy we determine the sleep and wakeup periods. For the MAC layer channel sensing, we use Clear Channel Assessment Mode -1 (CCA Mode 1); a simple energy detection scheme available as part of the IEEE 802.15.4 standard.

\section{Smart Application - The model}

We argue that node based task scheduling and pre-emption is perhaps not best suited for multihop EHWSN. For instance, it is possible that low priority tasks located at the head of the queue gets executed during energy stress disregarding network's requirement and energy budgets. We propose that applications have to be ``smart''. A typical WSN application comprises of operations such as sense, compute, store and communicate. These operations are required to be executed based on a policy over several time slots. Communication being an integral part of a node, it typically comprises of packet transmission and reception. Packet forwarding is essentially transmission but refers to relaying (reception and
transmission) a neighbour's packet. Transmission power control is mandated due to energy considerations. The smart application is essentially a set of policy elements. Smart application design and its successful implementation require several parameters:
(a)~Available energy in the storage buffer. We call this the
``\textit{real energy}'' ($E_A$). (b)~Predicted energy that would be harvested in the next two or three time slots called ``\textit{virtual energy}'' ($E_P$).
(c)~The ``\textit{network energy}'' ($E_N$) which is essentially the energy available in the neighbouring nodes (can be extended to energy available along a route). Thus our smart application has to exploit the energy in the slot to the maximum such that the ``\textit{residual energy}'' between the available energy and consumed energy in a slot is minimum. Since energy replenishment occurs in EHWSN, energy utilization should be such that maximum number of tasks or operations should be completed. As one can observe, Packet Reception Ratio (PRR) and application performance depends on accurate measurement of the above parameters. Since available energy is stored in a super-capacitor, we calculate the real energy by sensing the voltage across the capacitor. We use Holt-Winters model to predict the virtual energy in a node. We leverage on the RSSI values contained in routing messages for periodic update on the energy level in the network. Since available energy on each node varies, the application morphs itself into several forms in a manner that performance deterioration is within acceptable limits. Though our analysis and simulations can use many energy levels, for the purposes ease of implementation, we discretize the available energy levels into $4$ levels represented by $E_i, i \in \{0,1, 2, 3\}$. $E_{0}$, $E_{1}$, $E_{2}$ and $E_{3}$ correspond to \texttt{Lowest survivable}, \texttt{Minimum}, \texttt{Intermediate} and  \texttt{High} energy levels respectively. Similarly, the residual energy levels are represented by $E_i^\prime, i \in
\{0,1, 2, 3\}$, where $E_0^\prime$, $E_1^\prime$, and $E_2^\prime$,
$E_3^\prime$ correspond to \texttt {Lowest survivable, Minimum,
Intermediate and Higher} energy levels respectively. Let
the stored energy, harvested energy, predicted (virtual) energy at
time slot $k$ on node `$n$' be represented by $E_{S(k)(n)}$,
$E_{H(k)(n)}$ and $E_{P(k)(n)}$ respectively. Let $E_{N(k)(s)}$ be
the energy available on neighbouring node `$s$' at time slot $k$
and $E_{DC(k)(n)}$ denotes the minimum energy required for the
node for one duty cycle. We denote the maximum energy capacity of
a supercapacitor as $E_{max}$. The harvested energy for node $n$
is $E_{H(k)(n)} \ge 0$ and it is a trivial assumption as no energy harvested can only stall the application execution. The available energy at each time slot $k$ on node $n$ is represented by
$E_{A(k)(n)}$, $E_{A(k)(n)} = E_{S(k)(n)} + E_{H(k)(n)}$ and
$E_{A(t)(n)} \leq E_{max}$.

\subsection{The Base Application and Morphing}

As mentioned earlier, the base application essentially constitutes a set of policies which would be executed if available energy is high. At the beginning of each time slot, the application checks the available energy, virtual energy and the policy that requires to be executed. If there is sufficient harvested energy, the node initiates network support for its policy and prepares itself to execute the corresponding policy, else, the application on the node decides to morph into a slower version (lesser functionalities than the base application). In the event of network showing lack of support, e.g., neighboring nodes having lesser energy, the application will be \textit{again} forced to morph itself to a further slower version. For example, when ensuing two time slots reports the virtual energy for the node is low and current energy level is low, the application morphs to a lower version by partial execution of the policy with a few operations backlogged for later execution. A partial fulfillment for a specific policy is necessary to ensure that priority for certain operations in the policy has soft guarantees.
\begin{figure}[h]
\centering
\centering \includegraphics[scale=0.18]{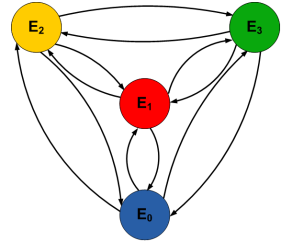} \caption{Energy
level transitions on each energy harvesting node. } \label{state}
\end{figure}

\begin{table}
\caption{Energy consumed/operation by our custom motes}
\label{table_energyperoperation} \fontsize{8}{8} \selectfont
\centering
\begin{tabular}{|c|c|} \hline
Operation & Energy Consumed\\ \hline
{Average of 50 samples} & 7.056 $\mu$J \\ %\hline
\hline\ {Finding peak among 50 samples} & 7.392 $\mu$J \\%\\hline
\hline\ {Sensing once from ADC} & 20.30 $\mu$J \\%\\hline
\hline\ {Writing 1 byte to Flash} & 1.23 $\mu$J \\%\\hline
\hline\ {Reading 1 byte from Flash} & 0.3 $\mu$J \\%\\hline
\hline\ {Transmitting 128 bytes @ 0dBm} & 0.341 mJ  \\
\hline\ {Receiving 128 bytes once} & 0.40 mJ\\ \hline
\end{tabular}
\end{table}
\normalsize 
Fig.\ref{state} captures the time varying energy profile for a node and shows the energy transitions based on real, virtual and network's energy. For instance, the state transition diagram shows when the node is high on own energy and virtual energy (i.e. both the energy values are at $E_3$), the node can complete its policy execution in the current time slot including any backlog operations. This energy expenditure is possible only when the network provides the necessary support. With its completion, the node can transit to either $E_1$ or $E_0$ depending on the level of energy depletion. At the same time, if the future time slots predict a low energy, the base application chooses to morph to a slower form and thus deplete lesser energy and transit to say either to $E_2$ or $E_1$.

\subsection{Policy Models}

\subsubsection{Policy Model - Operation Set:}

The operation set includes the set of operations (both node and network) any wireless sensor node performs. Basic operations may include sensing, computing, communication and storage. Each element in the operation set is associated with fixed energy consumption. The typical energy requirements for each operation on our custom mote are given in Table \ref{table_energyperoperation}.

\subsubsection{Policy Model - Policy Set (P):}

A policy set $P$ defines the set of operations and their order of execution obtained from an operation vector that a sensor node $n$ has to follow. Typically, a sensor node might be associated with two or more policies, and the specific set of policy to be executed in a particular time slot is defined as part of the base application. Thus the policy set for a node can be defined as,
\begin{equation}
 \label{policy} P =\left\{{\begin{array}{c}
 P_{1},
 P_{2},
 P_{3},
 .
 .
 .
 ,P_{x}
 \end{array}}\right\}.
\end{equation}
\normalsize 
A typical policy on a sensor node $D$ can be $P_d =\{P_{1}, P_{2}, P_{3}\}$ where, $P_{1}$ = \{Sense,
Transmit\}, $P_{2}$ = \{Sense, Compute, Write\} and $P_{3}$ =
\{Read, Compute, Transmit\}. For example, a relay node $R$ may have the policy set $P_r$ = \{$P_{4}$, $P_{5}$, $P_{6}$\} where, $P_{4}$ = \{Receiving, Forwarding\}, $P_{5}$ = \{Read, Compute, Forward\} and
$P_{6}$ = \{Receive, Write\}. At each time slot, the node decides to execute subset of policy set denoted by ${P^\prime}$. The energy required for
policy execution in time slot $k$ is given by, 

\begin{equation}
\label{policy-energy}
E_{R(k)} = \sum_{i=1}^{m}E_{i} {P_i}^\prime, ~ ~ ~ 
 \begin{cases}
m=x, P_i^\prime = P\\
m<x, P_i^\prime \subset P
\end{cases}
\end{equation}
\normalsize 
where, ${P^\prime}$ is the subset of the policy set,
$x$ the maximum number of policies and $E_i$ is the energy consumed for $i^{\mbox{th}}$ policy set. 

\section{Solar Energy Prediction Model}
\label{solarenergy}
One of the key requirements for optimal application performance is accurate prediction of virtual energy on each node. Virtual energy availability has impact not only on node's own performance but also on the network's performance. As mentioned earlier, literature on time series models for energy prediction is a well researched area. However, none till date have performance results from a physical implementation. In order to study network wide efficacy of time series models, we used solar energy trace database with data collected over three years by lowry range solar station \cite{lrss} from year May 2008 to August 2011. Three data sets are available from the website corresponds to direct sunlight, reflected sunlight, and diffused sunlight. Since the complete database of three years requires about $450$ KB for each dataset, it was easily possible to load the database on individual sensor nodes. We applied these spatio-temporal data sets across the complete network as shown in Fig \ref{setup}. We evaluated two of the most commonly used energy prediction models namely Exponentially Weighted Moving Average [EWMA] prediction model, and the Holt-Winters [HW] time series prediction model for these data sets.

\subsection{ EWMA Prediction Model}

EWMA is a commonly used algorithm for predicting solar energy which assigns weights to data in time series, assigning lower weights to older data and giving importance to more recently acquired data. The forecast value using the EWMA algorithm is obtained from Eq. \ref{eq:EWMA}.

\begin{equation}
\label{eq:EWMA}
E_{P}(t+1) = \epsilon E_{A} + (1- \epsilon) E_{P}(t)
\end{equation}

In the Eq. \ref{eq:EWMA}, $E_{P}(t+1)$ indicates the predicted value for the time slot $t+1$, $E_{A}$ indicates the measured available energy at time slot $t$ and $0$ $<$ $\epsilon$ $<$ $1$ is the weighting factor. To predict the incoming energy, EWMA sums the last read real value to the previous value predicted with weights $\epsilon$ and 1-$\epsilon$. If the weighing factor $\epsilon$ is high i.e. close to 1, previous values will have higher importance and vice versa. The best value for $\epsilon$ is choosen based on the least mean square error (LMSE). We found that $\epsilon$ value of 0.5 provides a minimum error for the EWMA algorithm which is the same as chosen in \cite{aman}.

\subsection{ Holt-Winters Prediction Model}

Since solar energy output can have seasonal variation, intelligent prediction algorithms should exploit the trend and seasonality available within the data sets. We applied Holt-winters time series prediction model as it is a commonly used time series model for forecasting when data follows a trend and seasonality. The Holt Winters algorithm computes forecast value based on the estimates of trend and seasonality from the previous data. HW algorithm uses the following set of equations for forecasting:

\begin{equation*}
\label{eq:HW}
E_{P}(t) = \epsilon \frac{y(t)}{I(t-l)} + (1-\epsilon) (E_{P}(t-1) + b(t-1))
\end{equation*}
\begin{equation*}
\label{eq:HWtrend}
b(t) = \gamma(E_{P}(t) - E_{P}(t-1)) + (1-\gamma)b(t-1)
\end{equation*}
\begin{equation*}
\label{eq:HWseasonal}
I(t) = \beta \frac{y(t)}{E_{P}(t)}+(1-\beta)I(t-l)
\end{equation*}
\normalsize

In the above equations, $E_{P}$ is the predicted value, $y$ is the current value, $\epsilon$ is the weighting factor, $b$ is the trend factor, $I$ is the seasonal index, $t$ is the current time slot, $l$ is the number of periods. The term $ \epsilon $, $ \beta $, $ \gamma $ are constants whose values are estimated such that their LMSE is minimized and estimated it to be $0.906$, $0.1$ and $0.650 $ respectively.

\section{Multi-criteria Optimization Problem}
\vspace{-5pt}
We mentioned in previous sections about efficient energy utilization in each time slot where the difference between available energy and consumed energy is minimized for a given real, virtual and network energy. The problem is to find an optimal policy execution. We cast this problem as multi-criteria optimization with two objectives: (a)~Maximizing application policy  execution, and (b)~Minimizing residual energy.
\vspace{-7pt}
\subsection {Maximizing application policy}
\vspace{-5pt}
In each time slot $k$, the objective is to maximize
$P^\prime$ policies executed by a node. Thus the application
policy utility $U$ is given by, 

\begin{equation}
\label{policy-utility}
U=  \sum_{i=1}^{m} \alpha_{i} {P_i}^\prime  ~~~~~ s.t. 0 \leq
\alpha_i \leq 1
\end{equation}
%\begin{equation*}
%\label{max-policy-1}
%s.t. ~~~~~  0 \leq \alpha_i \leq 1 %~~  \sum_{i=1}^{m} {P_i}^\prime \leq P_{x}, ~~ m \leq x;
%\end{equation*}
\normalsize
In Eq.(\ref{policy-utility}), ${P_i}^\prime$ is the set of policies to be executed and $\alpha_i$ indicates the morphing factor for the $i^{\mbox{th}}$ policy set. When $\alpha_i$ = 1, the entire policy set $P$ is executed (i.e., ${P_i}^\prime = P$).
\vspace{-7pt}
\subsection {Minimizing residual energy}
\vspace{-5pt}
Let $X_k$ be the total available energy in time slot
$k$. This is a function of $E_{A(k)}$, $E_{P(k)}$  and $E_{N(k)}$.
Let $Y_k$ be the energy required to execute the policies in time
slot $k$ as given by Eq. \ref{policy-energy} and $Z_k$ is the energy required for operation of the node with a duty cycle factor of $\delta$. In the event of $X_k$ $<$[$Y_k - Z_k$], we have 

\begin{equation}
\label{y-optimal}
\hat{Y_k} =  f(Y_k, \alpha_k) ~~~~~ s.t. ~~ 0 \leq
\alpha_k \leq 1;
\end{equation}
\begin{equation}
\label{z-optimal}
{Z_k} = f( E_{DC}, \delta_k) ~~~~~ s.t. ~~ 0 \leq \delta_k \leq 1;
 \end{equation}
\normalsize
$\hat{Y_k}$ is the consumed energy to execute the set
of policies based on the morphing factor `$\alpha$' and $Z_k$ is the energy required for adaptive duty
cycle based on harvested energy in each time slot. We define the
residual energy utility, $V$, as 
\begin{equation}
\label{energy-utility} V=    [X_k - \hat{Y_k} - {Z_k}]
%\sum_{i=1}^{m}
\end{equation}

\begin{equation*}
\label{energy-utility-1} s.t. ~~~~~  X_k \leq E_{max}, ~~ [X_k -
\hat{Y_k} - {Z_k}] \geq E_0^\prime ;
\end{equation*}
\normalsize where $E_0^\prime$ is the lowest survivable residual energy level required for
the system to be operational. Now, our multi-criteria optimization
problem is formulated as, 
\begin{equation}
\label{mopt-problem} 
~~~\mbox{MOPT:} ~max ~ U= ~\sum_{i=1}^{m}
\alpha_{i} * {P_i}^\prime
\end{equation}

\begin{equation*}
\mbox{and}~~~~~~~~~~~~~min ~ V= ~  [X_k - \hat{Y_k} - {Z_k}]
%\sum_{i=1}^{m}
\end{equation*}
\normalsize First we consider a single objective optimization
problem for a given $V$ (i.e., fixing one of the objectives).

\begin{equation}
\label{opt-problem} \mbox{OPT}(V)~max ~~ U= ~\sum_{i=1}^{m}
\alpha_{i} * {P_i}^\prime;  ~~~~s.t. ~  [X_k - \hat{Y_k} - {Z_k}] = V
\end{equation}
%\begin{equation*}
%s.t. ~~~~~~~  [X_k - \hat{Y_k} - {Z_k}] = V
%\sum_{i=1}^{m}
%\end{equation*}
%
For all possible values of $V \in [E_0^\prime, E_{3}^\prime]$ we obtain their
corresponding $\mbox{OPT}(V)$, optimal points in application
policy versus residual energy curve. This gives a mapping from $V$
to $U$, which is denoted as $f: V \rightarrow U$, where for each
point $(U,V), U=f(V)$ is the maximum application policy utility
that can be obtained.
\vspace{-5pt}
\section{Solution}
\vspace{-5pt}
The solution to $\mbox{OPT}(V)$ is obtained by using
the special structure of linear programming such as the Parametric
Analysis (PA).  The overall approach is to obtain $f(V)$ by
solving a finite number of linear programs to provide the morphing
factor ``$\alpha$'', transmission power factor and duty cycle
factor ``$\delta$''.  These values ensure minimization of residual
energy and maximization of application policies.

Using PA, we study the perturbation of $V$ and its effect on the
optimality of $\mbox{OPT}(V)$ to obtain the perturbation factor
$\lambda$. We use boldface to denote matrices and vectors. For a
given $V$, the current optimal basis of $\mbox{OPT}(V)$ could
still be optimal when there is a perturbation on $V$. Thus, the
range $[E_0^\prime, E_{3}^\prime]$ can be partitioned into consecutive small
intervals, each corresponding to a different optimal basis. 

We use Dual simplex algorithm to find the optimal basis in each iteration. The dual simplex method is useful for re-optimizing a problem after a constarint has been added or some parameters have been changed so that the previous optimal basis is no longer feasible.

The algorithm starts with a basic solution that is dual feasible so all the elements of `row 0' must be nonnegative. the iterative step of the algortihm first finds the variable that must leave the basis and then finds the variable that must enter the basis to maintian dual feasibility.

For a
given $V$, let the optimal basis matrix be $\mathbf{B}$ and the
non-basic matrix be $\mathbf{N}$. Let the optimal solution to
$\mbox{OPT}(V)$ is $(\mathbf{x_B},\mathbf{x_N})$, where $x_B$ and
$x_N$ denote the values of basic and non-basic variables
respectively. Further, let $\mathbf{c_B}$ and $\mathbf{c_N}$
denote the coefficient vectors of the objective function of
application policy utility $U$ for the basic and non-basic
variables respectively. $\mathbf{b}$ is the vector with
coefficients of $V$ . The corresponding canonical equations are:

\begin{equation*}
U + \mathbf{(c_B B^{-1} N -c_N)x_N=c_BB^{-1}b }
\end{equation*}

\begin{equation*}
\mathbf{x_B +  B^{-1} N x_N=B^{-1}b}
\end{equation*}
 \normalsize

Let the perturbation on parameter $V$ be $V+\lambda$.  Then the
vector $\mathbf{b}$ is replaced by $\mathbf{b}+\lambda
\mathbf{b^{'}}$ with vector $\mathbf{c_BB^{-1}N-c_N}$ unchanged.
Now  $\mathbf{B^{-1}b}$ will be replaced  by $\mathbf{B^{-1}
(b+}\lambda \mathbf{b^{'}})$ and accordingly the objective becomes
$\mathbf{c_B B^{-1}(b+}\lambda\mathbf{ b^{'}})$. As long as
$\mathbf{B^{-1} (b+}\lambda \mathbf{b^{'}})$ is non-negative, the
current basis remains the optimal basis. The value of $\lambda$
for another basis to become optimal can be determined as follows:
Let $S=\{i: \overline{\mathbf{{b_i}^{'}}} < 0\}$ where
$\overline{\mathbf{{b_i}^{'}}} = \mathbf{B^{-1}b^{'}}$. If
$S=\emptyset$, then the current basis is optimal for all values of
$\lambda \geq 0$. Otherwise, let

\begin{equation}
\label{eq:basiseq}
\hat{\lambda} = \min_{i\in S}
\frac{\overline{{\mathbf{b_i}}}}{-\overline{\mathbf{{b_i}^{'}}}}.
\end{equation}
\normalsize
Let $\lambda_1=\hat{\lambda}$, then the current
optimal basis is optimal for $\lambda \in [0,\lambda_1]$, where
$\mathbf{x_B = B^{-1} (b+}\lambda \mathbf{b^{'}})$ and the optimal
objective is $\mathbf{c_B B^{-1}(b+}\lambda \mathbf{b^{'}})$. When
$\lambda>\lambda_1$, the basis $\mathbf{B}$ is no longer optimal.
Thus, we need to choose a variable $\mathbf{x_r}$ to leave the
basis, where the minimum in Eq.(\ref{eq:basiseq}) is attained for
$i=r$. $\mathbf{x_s}$ is chosen by the dual simplex method
rule~\cite{linearprog} and we update the canonical equations based
on the new optimal basis obtained and get $(U,V)$ pair as defined
earlier for $\mbox{OPT}(V)$. The process is repeated to find the range $[\lambda_1,\lambda_2]$
over which the new basis is optimal.

\begin{algorithm}
\caption{Basis Updation Algorithm using Dual simplex method}
\label{algo1}
\begin{algorithmic}[1]
\STATE
\textbf{Input:} An optimal basis matrix $\mathbf{B}$ for a given V.
\STATE
Compute $\overline{\mathbf{N}}$ = $\mathbf{B}^{-1} \mathbf{N}$, $\overline{\mathbf{b}}$= $\mathbf{B}^{-1}\mathbf{b}$ and $\overline{\mathbf{{b_i}^{'}}}$ = $\mathbf{B}^{-1}\mathbf{I}$ \\
\STATE
If S = $\{$ i : $\overline{\mathbf{{b_i}^{'}}}$ $<$ 0 $\}$ = $\emptyset$ terminates.\\
\STATE
$\hat{\lambda} = \min_{i\in S}
\{\frac{\overline{{\mathbf{b_i}}}}{-\overline{\mathbf{{b_i}^{'}}}}\} $\\
\STATE
r = arg $min_i \{\frac{\overline{{\mathbf{b_i}}}}{-\overline{\mathbf{{b_i}^{'}}}}\}$\\
\STATE
s= arg $min_j \{\frac{\overline{{\mathbf{b_j}}}}{\overline{\mathbf{N}}_{jr}}, \overline{\mathbf{N}}_{jr} < 0\} $\\
\STATE
Let $new (\mathbf{B}) = ( new(B) /\ {r} )$ $ \cup$  ${s} $ and $\mathbf{b} = \mathbf{b} + \hat{\lambda} \mathbf{I}$.\\
\STATE
Update $\mathbf{B}$ based on $new\mathbf{(B)}$.\\
\STATE
Compute V = V + $\hat{\lambda} $ and $U = c_B\mathbf{B^{-1}b} $\\
\STATE
\textbf{Output:} The new basis $new \mathbf{(B)}$, $\hat{\lambda}$ and (U,V) pair.\\
\end{algorithmic}
\end{algorithm}

Algorithm \ref{algo1} descibes the steps to obtain the new optimal basis for a given basis matrix $\mathbf{B}$. Thus, starting from $V=E_0^\prime$,
we repeat the steps iteratively to find different bases until we
reach $E_{3}^\prime$. The series of $\hat{\lambda}$ for these bases
will partition $[E_0^\prime, E_{3}^\prime]$ into small intervals. Thus, by
executing the above steps repeatedly, we obtain a series of
$(U,V)$ pairs, each corresponding to an optimal basis. We obtain
the application policy versus residual energy optimal curve by
connecting these endpoints consecutively.

Thus for each small interval with an optimal basis, we now show that f(V) is linear. Suppose, interval  $[E_0^\prime, E_{3}^\prime]$  is divided into 'K' small intervals [ $E_i^\prime$, $E_{i+1}^\prime$], i= 1,...,K, where  $E_1^\prime$ = $E_0^\prime$, $E_{K+1}^\prime$ = $E_3^\prime$ and the optimal basis for each small interval [ $E_i^\prime$, $E_{i+1}^\prime$] is $\mathbf{B_i}$. Then, the objective value function f(V) can be computesd as 

\begin{equation*}
f(V) = \mathbf{c_B B^{-1}(b+}\lambda \mathbf{b^{'}})
\end{equation*}

where $\lambda$ = $E_0^\prime$ - $E_i^\prime$
Thus,

\begin{equation}
\label{linear}
f(V) = \mathbf{c_B B^{-1}(b+} ( E_0^\prime - E_i^\prime) \mathbf{b^{'}})
\end{equation}

In Eq \ref{linear}. $c_B$, $\mathbf{B^{-1}}$, b, I are constants and $E_i^\prime$ is the only variable and hence f(V) is a linear function.

\begin{figure}
\centering
\includegraphics[scale=0.3]{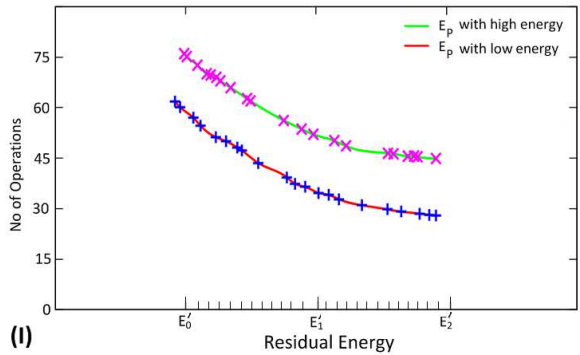}%e3_diff.eps}
\caption{ Optimal application policy versus residual energy when
available energy is $E_3$} \label{curve-e3}
\end{figure}

\begin{figure}
\centering
\includegraphics[scale=0.25]{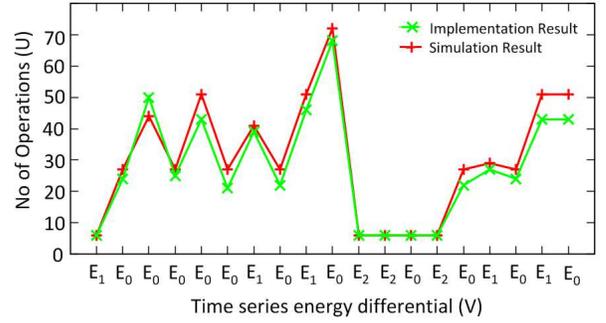}%{impvssim.eps}
\caption{  Plot of time series residual
energy versus number of operations.} \label{curve-e3}
\end{figure}

Fig.\ref{curve-e3}(I) shows the simulation results where x-axis
represents the residual energy (V) between the available and
the consumed energy and y-axis represents the total
number of operations executed by the node i.e., $U$.
Fig.\ref{curve-e3}(I) shows the optimal curve for a node when
available energy is $E_3$ and its neighbour node energy is either
$E_1$ or $E_2$ or $E_3$. Curves 1 and 2 indicate the optimal curve
when the energy prediction for future slots is high and low
respectively. When the available energy on the node was $E_3$, the
morphing factor $\alpha$ obtained were $0.9890$, $0.6269$,
$0.2797$ for residual energy of $ E_0^\prime $, $ E_1^\prime $ and
$ E_2^\prime $ respectively. Similar curves are obtained for other
available energy levels i.e., $E_2$, $E_1$ on the source node.
Fig.\ref{curve-e3}(II) shows the time series of residual energy
against the number of operations performed by the source node
obtained via simulation and implementation. The source node runs the diffused energy profile from the LRSS. We can see that. 
the simulation result closely matches the implementation. The
difference in number of operations between simulation and
implementation is around 6\% due to error in prediction. For
instance, at time slot $3$ of Fig.\ref{curve-e3}(II), the
available energy at the source node was wrongly predicted and
hence the node performed more operations compared to the
simulation result.

\section{Network and MAC Layer protocol Enhancements for EHS nodes}
In EHS, because each node in the network can harvest energy at various rates, it is necessary to use energy harvesting aware mechanisms at all layers in the protocol stack. Also, as our smart application is expected to run on a network of sensor nodes, in this section we list the key objectives of ``energy harvesting aware'' network and MAC layer. Furthermore, since routing and MAC layers are well researched in literature, our approach is to identify candidate routing and mac protocols and perform necessary modifications to meet our objectives. Our main focus is to design and develop a multi-node multi-hop energy harvested sensor nodes for outdoor applications like intrusion detection, smart buildings etc.

\subsection{Network Layer Enhancements}
Given a network wide varying energy profile, one major challenge in multi-hop energy harvested WSNs is to find stable routes from the source node to the sink (or base station) over a set of communication nodes. In a typical scenario shown in Fig \ref{setup}, given a set of possible routes, the question is to find the best route from node D to node A.  The selected transmission power at the source and relay nodes should ensure that receiver gain is sufficiently high such that it overcomes channel impairments such as fading. Due to this reason, transmission power control is one of the key requirements.

\begin{figure}[ht]
\centering
\centering \includegraphics[scale=0.42]{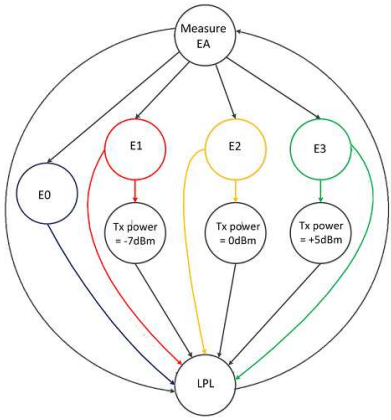}
\caption{Selection of transmission powers based on the available energy. }
\label{state}
\end{figure}

Since we have chosen $3$ discrete energy levels, we have divided the supported radio transmission powers also into $3$ discrete values as shown in the figure \ref{state}. We use this mapping in the multi criteria optimization setting as well. Due to energy fluctuations, an on-demand routing protocol is required to ensure route stability inspite of nodes abruptly shutting down and powering up. We selected AODVjr, a reactive routing protocol as a candidate. In the beginning, EHS nodes send out a beacon packet using a fixed transmission power.  Relay nodes that listen to this beacon use it as a reference received power.  The beacons are sent out at regular intervals to ensure that the reference is constantly adjusted to the environment and channel. From our measurements, RSSI reference level of -$65$ dBm was observed for a transmission power of $0$ dBm at a typical inter-node distance of about $25-30$ feet. Thus each relay node is apriori aware of the signal strength of packet reception from their downstream relays in the network. To setup routes, source nodes transmit route request packets called the RREQ broadcast packets using a transmission power supported by its available energy. This procedure is followed at each relay node that forwards the received RREQ from the source node towards the base station.

We develop a modified AODVjr protocol for EHWSN. In our Modified-AODVjr case, the sink node performs the weighted sum of the link quality against each candidate route and replies with an RREP packet that carries information about the best route back to the source. The best route is the one between the source and destination that has the least weighted sum of the link qualities obtained from Eq. \ref{eq:costrssi}
\begin{equation}
\label{eq:costrssi}
C_{(s,d)} = \sum\limits_{i=1}^n \mu_i * RSSI_i
\end{equation}

In  Eq. \ref{eq:costrssi}, $C_{(s,d)}$ indicates the cost from source node to the destination node, $RSSI_i$ indicates the Received Signal Strength Indicator (RSSI) of node 'i' and $ 0 < \mu \le 1 $ is the weight assigned based on the RSSI value at each hop-node in the path from source to destination node. The term $\mu$ is obtained from extensive measurements performed on our custom board by varying both transmission powers and inter-node distance.

Since the weighted sum at the sink node is dependant on the network wide energy fluctuations, another modification to AODVjr facilitates every RREQ message to be honoured even in the middle of a data transfer. The latest RREQ messages are used to recompute the weighted sum and if this sum is found to be lesser compared to the previously computed value, the same is conveyed back to the source node via a RREP message. Unlike AODVjr, we reintroduced source sequence numbers to RREQ messages to improve the route reliability. This modification also takes care of dropped and lost RREP packets. Thus Modified-AODVjr compares both the sequence number and overall link quality of the route before sending an RREP.

\subsection{MAC Layer Enhancements}

We believe ``Virtual Energy Transfer'' between peer nodes is an important objective for energy harvesting aware MAC layers.  This  comes in two related ways: (a) In energy harvested systems, since harvesting rate is dependent on the environmental conditions, the node has to optimize its sleep and wakeup schedules based on the harvested or available energy. For instance, the idle (reception) current for CC2520 is $22.3$ mA. Thus, one of the important methods to reduce the energy consumption of the device is by duty cycling. However, an energy harvesting node high on energy would perhaps increase its duty cycle to support and respond to shorter preambles; thereby reducing the energy consumed by the sender. (b) Choice of transmission power for control and data packets should also be based on available energy.  In CC2520, the transmission (0dBm) and reception currents are $25.8$ mA \& $18.5$ mA respectively. A node high on energy can include table entries of other high energy nodes; thereby skip entries for low energy neighbours.

We perform enhancements on one of the well known and popular MAC protocols called the BOXMAC-2 protocol \cite{boxmac}. BOXMAC-2 is a random access MAC protocol and uses LPL with periodic check times for ongoing communication. To reduce overhearing between neighbours, a long preamble is replaced by a short preamble that consists of packets containing the receivers address along with a small receive check period between successive packets. Hence, packet-based MAC protocol is well suited for harvesting systems due to the fact that it does not require any initial arbitration and only the sender and receiver pair spends energy. We have used its ability to draw cross-layer information such as the term $\delta$ to adapt the residual energy to the duty cycle.

\[ \delta =
\begin{cases}
0 & ~~ Only~ system ~related ~operations\\
Otherwise & ~~ energy ~transfer ~is ~enabled
\end{cases}
\]

In the event of  $\delta$ taking a value  $0$, the node does only system related operations as there is no energy for communication.  For all other values of $\delta$ the virtual energy is enabled with maximum transfer to neighbours at $\delta$ = $1$.

\section{Results}
In this section, we discuss the results pertaining to: (a) Efficacy of energy prediction models, (b) Routing layer performance for varying energy profiles, and (c) Adaptive duty cycling on EHS nodes based on the harvested energy.

\subsection{Evaluation of energy prediction models}

\begin{figure}[htp]
\centering
\centering \includegraphics[scale=0.28]{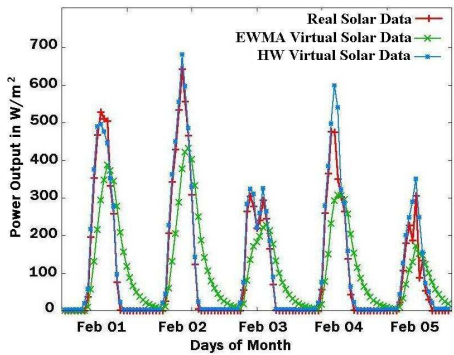}
\caption{Comparison of real, EWMA and HW solar energy}
\label{ewmaplot}
\end{figure}

We compare the real solar data with virtual solar data obtained from EWMA and HW prediction models for five days of a month. The first, second and the fourth days correspond to sunny conditions and the second, fifth day represents the cloudy situations as shown in Fig  \ref{ewmaplot}. Since EWMA prediction algorithm only uses values from previous days at the same time period, if the weather condition changes for the next day, this method presents a large error in prediction. Looking at the Fig. \ref{ewmaplot} it can be easily noticed that EWMA algorithm fails to adapt to changes in weather on consecutive days. The HW model computes the terms $\epsilon$, $\beta$, $\gamma$ based on the previous solar data to obtain the weighting, trend and seasonal factors. It can be clearly seen from the Fig. \ref{ewmaplot}, HW prediction model has the ability to predict the solar values by considering both seasonal and weather condition changes from one day to another. This improvement in prediction of solar energy comes from the trend and seasonal factors considered by the prediction model. From the three year datasets, we evaluated the maximum average error percentage of about 45\% for EWMA prediction model and around 7\% for Holt-Winters energy prediction model.

\subsection{Routing Layer}

\begin{figure}[htp]
\centering
\centering \includegraphics[scale=0.15]{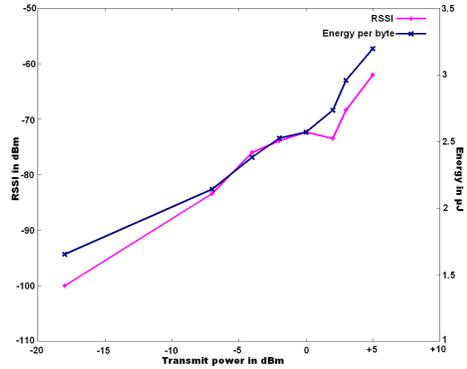}
\caption{Change in RSSI with transmit power}
\label{rssitx}
\end{figure}

\begin{figure}[htp]
\centering
\centering \includegraphics[scale=0.28]{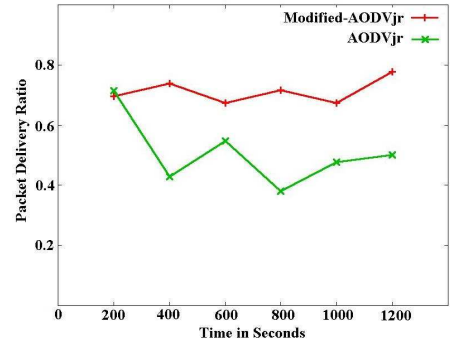}
\caption{Packet Delivery Ratio comparison for Aodvjr and modified Aodvjr protocols }
\label{pdr}
\end{figure}

Fig \ref{rssitx} shows the improvement of RSSI for increasing transmission powers. Data packets of $128$ bytes were transmitted every $2$ seconds. The figure also indicates energy per byte required for transmission on a node with various transmission powers. Thus, increasing transmission power requires higher energy per byte and hence node maps transmission powers to the corresponding energy levels. It is also noted from \cite{betafactor} that the channel coherence time is very high and hence change in RSSI is mainly due to change in transmission power. It is clear from the figure, transmission power change impacts RSSI and hence node changes the transmission power based on the energy level which is reflected on the RSSI value. We use this RSSI value to identify the neighbour nodes energy level and we call this as ``Faking of RSSI'' based on the energy level.
Our Modified-AODVjr routing protocol decides the best route to the destination node based on the weighted sum of the link quality. To study the performance comparison between AODVjr and Modified-AODVjr, we implemented both the protocols on our custom mote. Our metric for comparison is the packet delivery ratio. The experimental setup is as shown in Fig. \ref{setup}. Once the source node is powered up, it sends RREQ message to all its neighbours with the destination node replying either to the first RREQ (AODVjr) or reply based on the overall link quality of the route (Modified-AODVjr). In both the experiments the source node generates a data packet every $2$ seconds with a payload of $22$ bytes. Fig. \ref{pdr} shows the comparison of packet delivery ratio between AODVjr and Modified AODVjr. The Modified-AODVjr protocol has a average packet delivery ratio of $72.7$\% whereas AODVjr has an average packet delivery ratio of $51.2$ \%. Thus, Modified-AODVjr routing protocol is necessary for energy harvested systems to improve packet delivery ratio and quick route convergence.

\subsection{Adaptive Duty Cycling}
\begin{figure}[ht]
\centering
\centering \includegraphics[scale=0.22]{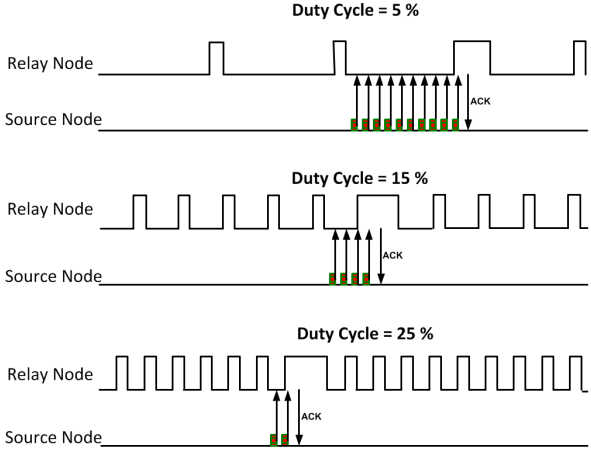}
\caption{ Adaptive Duty Cycle on EHS nodes based on harvested energy.}
\label{adaptive}
\end{figure}

As described previously, ``Virtual Energy Transfer'' between peer nodes is an important objective for energy harvesting aware MAC. We utilize the optimal $\delta$ value obtained from our optimization to set the duty cycle on EHS nodes. Fig. \ref{adaptive} shows three cases based on the $\delta$ value (a) The $\delta$ value of $0.05$ corresponds to duty cycle of ($5\%$) due to low energy available on the node (b) duty cycle of $15\%$ was chosen when the node has intermediate energy level and the obtained $\delta$ value is $0.15$, thus decreasing the sleep periods (c) when the harvested energy level is $E_3$, $\delta$ obtained is $0.25$, hence the duty cycle on the node is set to $25\%$.

In Fig. \ref{adaptive}, the source node has to send short packets as preamble for the duration of sleep time i.e., $100ms$ for $(5\%)$ duty cycle when the energy level is $E_1$. When harvested energy level on the node is higher i.e $E_2$, $E_3$ the number of packets sent as preamble by the source node to wakeup the neighbour should span the sleep interval of $35ms$ and $20ms$ respectively. Thus, reducing the number of short preambles sent by the sender and the energy spent by the source node when neighbour node is high on energy. Thus, enabling ``Virtual Energy Transfer'' between peer nodes at the MAC layer.

\subsection{Application}

\begin{figure}
\centering
\includegraphics[scale=0.20]{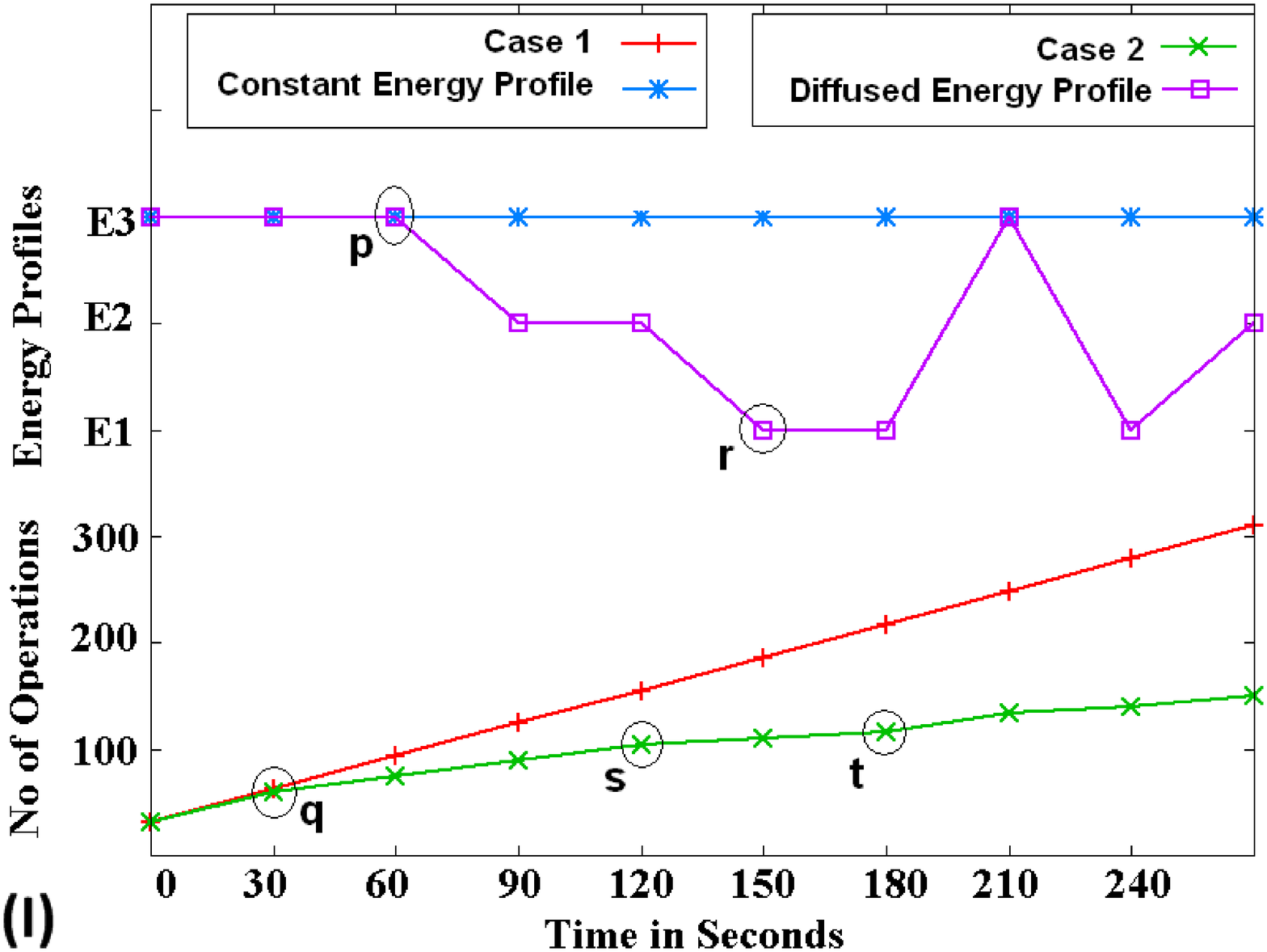}
%\caption{Behaviour of the morphed application on the source node for diffused and constant energy profile.}
\includegraphics[scale=0.20]{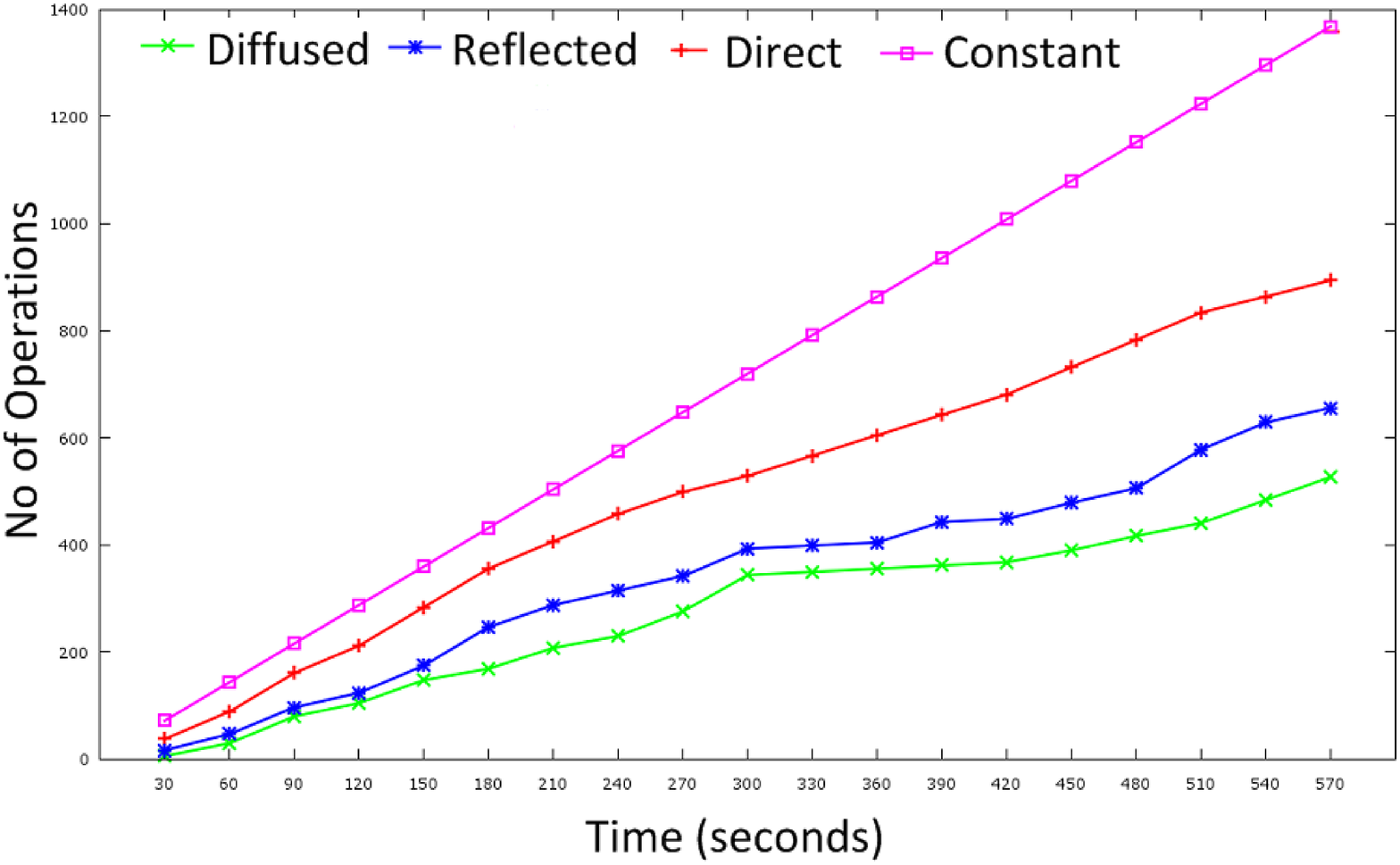}
\caption{(I) Morphed application on the source node for diffused
and constant energy profile. (II) Morphed application on the source
node for various energy profiles.} \label{results}
\end{figure}

We show the performance of the distributed smart
application running on sensor Node D (in Fig.\ref{setup})
associated with a diffused energy profile. The
solar emulator placed above Node D generates the diffused energy
profile where most of the time either $(E_1)$ or $(E_2)$ amount of
harvested energy is available. The sensor node uses the
Holt-Winters energy prediction algorithm in every time slot to
obtain the virtual energy. For generating a reference, we exposed
Node D to a constant high energy profile to fix $E_3$.
Fig.\ref{results}(I) shows the behaviour of the morphed
application on sensor Node D. The plot shows the comparison for
two cases: \textit{Case-1} is the reference $(E_3)$ energy level
and \textit{Case-2} when the node is subjected to the diffused
energy profile.
%\begin{figure}[ht]
%\centering
%%\epsfig{file=./EPS/DSA_app.eps, height=2.0in, width=2.5in,}
%\centering \includegraphics[scale=0.15]{DSA_app.eps}
%\caption{Behaviour of the morphed application on the source node for diffused and constant energy profile.}
%\label{results}
%\end{figure}

In \textit{Case-1}, the source and the neighbour's energy was
always high at $E_3$ and thus Node D could perform all the
policies associated with the energy level $E_3$. This can be
considered as the best case and serve as a reference for the
distributed smart application. In \textit{Case-2}, the source is
associated with diffused energy profile and the neighbour node's
energy profile is reflected. Fig.\ref{results}(I) shows the
results of application morphing under energy fluctuation. We
captured these fluctuations as events and enumerated them as:
$(a)$ when the energy of node is low and predicted energy for
future time slots is high and vice-versa; $(b)$ when energy level
of the node is high but the network's energy level is low and
vice-versa. In Fig.\ref{results}(I), several regions namely
$p,q,r,s,t$ show energy level transitions as well as application
morphing. The region `$p$' indicates the change in energy level
from $E_3$ to $E_2$ on the source node. The region `$q$' indicates
the morphing of the application in one time slot before the change
in energy level on the source node. The application morphed to a
slower version of the base application as shown in region `$q$'
where the number of operations performed by the source node was
significantly less compared to the previous time slot. This
morphing improves the available energy on the energy buffer. The
region `$r$' shows the low energy profile $E_1$ on the source node
and finally region `$s$' shows the application morphing to a
slower version of the base application. This application morphing
event is due to source node energy level being low and future time
slot energy level is high. However, in the next time slot it can
be clearly seen that Node D morphed to a faster version towards
the base application as the future energy level is $E_3$ and
region `$t$' indicates the increase in number of operations
performed by the sensor Node D.
Fig.\ref{results}(II) shows the implementation result of
application morphing on the source node for various energy
profiles obtained from LRSS. The neighbouring node was running
reflected energy profile. As expected, it can be clearly seen that
application morphing was enabled even when the source node was
running on direct energy profile. In summary, the upper bound on
application performance is dependant on source node's own energy
as well as network's energy. Application performance when source
node runs other energy profiles is also shown.

\begin{figure}
\centering
\includegraphics[scale=0.25]{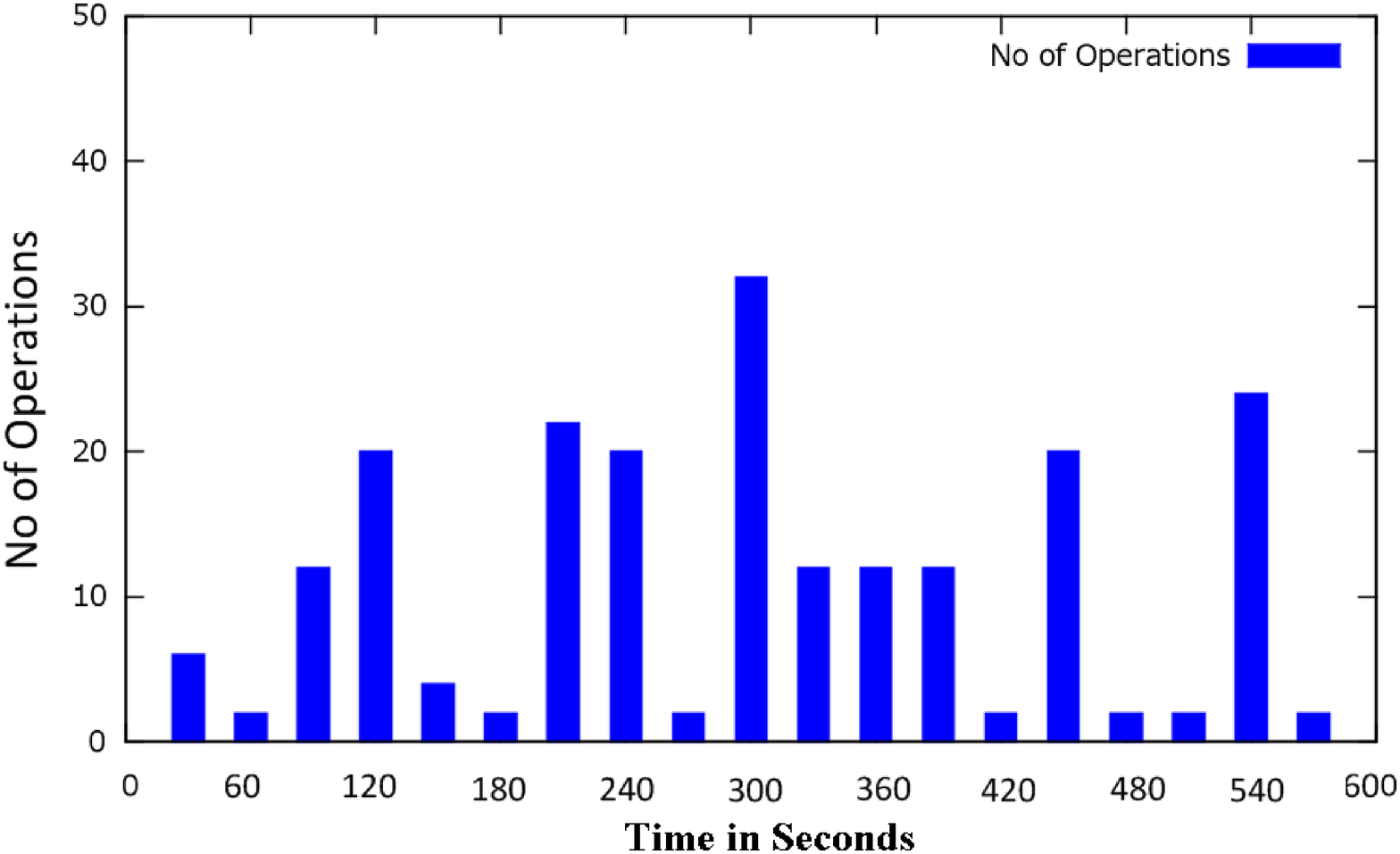}
%\caption{Behaviour of the morphed application on the source node for diffused and constant energy profile.}
\includegraphics[scale=0.25]{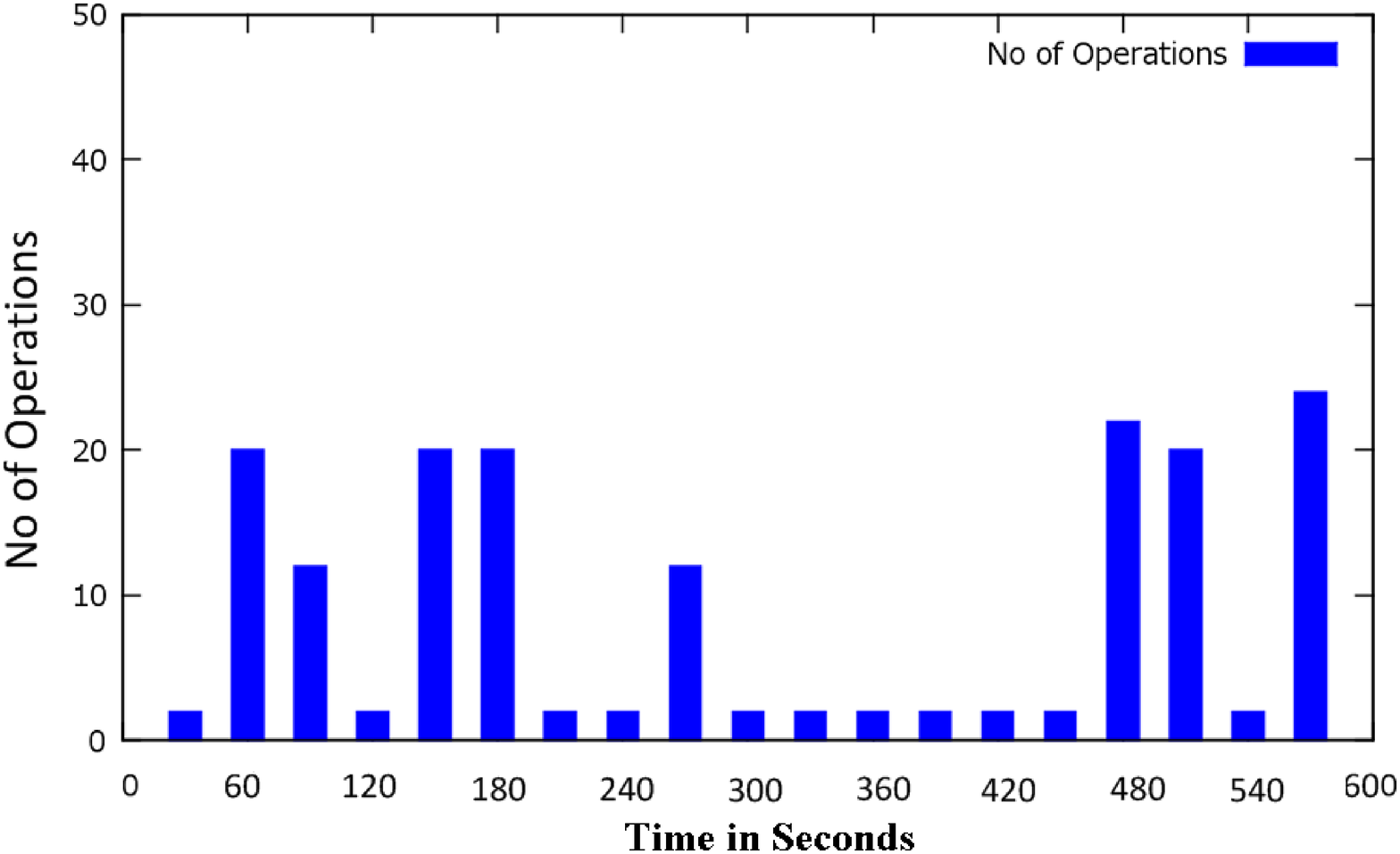}
\caption{(I) Number of operations performed by relay node B. (II)Number of operations performed by relay node C.} \label{relay}
\end{figure}

\begin{figure}
\centering
\includegraphics[scale=0.25]{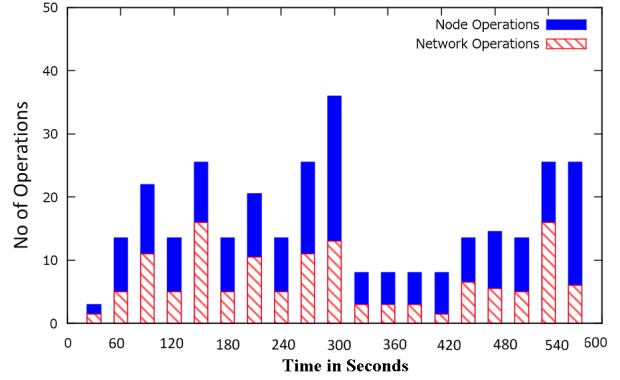}
\caption{(I) Number of node own and network operations performed by source node B.} \label{source}
\end{figure}

Fig. \ref{relay} shows the number of operations performed by the relay nodes B and C in each time slot. The relay node B runs direct energy profile and relay node C runs reflected energy profile. As can be observed from the figure, relay node B supported the source node to perform network related operations in most of the time slots. Figure \ref{source} shows the number of operations performed by the source node with diffused energy profile. The figure shows the number of node's own operations and network operations performed. In time slot $9$ the node was in intermediate energy level and the network returned intermediate support and hence, the source performed few network and its own operations. However, in the next time slot source node was in high energy level and network returned intermediate support, hence node performed less network operations and more node's own operations.

To study the efficacy of the optimization method, we enumerated number of node and network operations (refer Table \ref{table_energyperoperation}) performed by each node for all energy profiles. When the source node has diffused energy profile and neighbour has reflected profile, the source performed $354$ node operations and $172$ network operations. The source node performed $475$, $656$ node operations and $190$, $212$ network operations for reflected and directed energy profiles respectively. 

\section{Discussions }

In this paper, we discuss energy harvesting aware routing and MAC protocols. The prediction of virtual energy plays an important role on the application performance. As shown in section \ref{solarenergy}, we evaluated two of the most common time series prediction models. It can be clearly seen from Fig. \ref{ewmaplot}, if the real energy level was $E_3$, EWMA model forecasted the energy level as $E_2$ whereas HW model forecasted it as $E_3$ and when real energy level was $E_2$, the forecasted values were $E_1$ and $E_2$ by EWMA and HW models respectively. From our datasets, we evaluated the maximum error percentage of about $45\%$ for EWMA prediction model. Whereas, for HW model it is around $7$\%. Thus EWMA model based predictions deteriorate the application performance. In our experiments we found that when HW prediction model was used, on an average $64.5$\% of the operations were performed at the source node. The values of the parameters $\epsilon$, $\beta$ and $\gamma$ used in HW prediction model vary for different solar power traces. Concerning the routing protocol, significant energy savings were seen due to absence of control messages such as Hello, RERR etc. While the base station uses weighted sum of the link quality metric to provide a route, in reality, since power control is used by nodes, the link quality is a direct manifestation of the available energy. In the MAC layer, we believe ``virtual energy transfer'' ensures energy neutral operations for a multi-hop multi-node network setting.

\section{Conclusions}
We have implemented a multihop EHWSN using solar
energy. The efficacy of the proposed optimization algorithm was
studied by evaluating node and network operations for various
energy profiles. Clearly, the predicted, network and available
energy contribute to the working of the distributed application.
The smart application was able to maximize its operations by
adjusting its application policy (rate) to satisfy the least
residual energy criteria. In this work the number of hops is
limited and we have only showed possible way of building
application morphing also considering available energy in the
nodes of the network rather than source node alone. We are encouraged by
the results and propose to extend the network to include more
number of hops and entire routes to study the scalability of our scheme.  
To the best of our knowledge this is the first implementation of a multihop energy harvested WSN, \textit{albeit} two hops. We believe that our results are a step forward towards larger EHWSN  deployments. We plan to generalize this work by proposing a cognitive networking stack for EHWSNs. The idea here is to provide hooks that consider EH including predictions at every layer of the networking stack. Moreover, we propose to study the distributed algorithms also with handles such as dynamically varying available voltage and possible frequency scaling.

% that's all folks
\end{document}